# Critical temperature modification of low dimensional superconductors by spin doping

Pasi Jalkanen*, Vladimir Tuboltsev, Ari Virtanen, and Konstantin Arutyunov
Department of Physics, University of Jyväskylä, P.O. Box 35, FIN-40014 Jyväskylä, Finland

Jyrki Räisänen
Accelerator Laboratory, University of Helsinki, P.O. Box 43, FIN-00014 Helsinki, Finland

Oleg Lebedev and Gustaaf Van Tendeloo
Electron Microscopy for Material Science (EMAT, Physics Department, University of Antwerp, B2020 Antwerpen, Belgium

*) Fax: +358 14 260 2351, Electronic mail pasi.jalkanen@phys.jyu.fi

Interest to utilization of ion beams for study of properties of low dimensional structures is coming from the significant progress of ion implantation technique. Being certainly not applicable to bulk systems, it however provides two important advantages. First, ion implantation is capable of preparation of solid mixtures unobtainable by other methods. Second, typically targets for thin film deposition are prepared by melting mixtures of materials. Even if the resulting thin film is chemically stable, homogeneity might appear far from being perfect due to difference of deposition rates of the ingredients. Ion beam implantation in thin films can provide much better homogeneity, which later can be controlled by TEM technique. Sequential ion beam implantation enables also a study of evolution of impurity-host interactions eliminating inevitable artifacts of individual samples.

Here we report experimental results on Al film implantation with Fe and Mn ions. Concentration of dopants was varied to find out how superconducting properties of the metal can be modified at will. The purpose of Al film ion implantation was twofold. The first motivation was to study the basic physics of superconductivity in low dimensional metallic structures doped with magnetic impurities, so called spin doping. The second aim was to use magnetic ion implantation to suppress superconductivity for certain micro- and nanodevice applications, where this property is undesirable. A representative example is Coulomb blockage thermometry.

Interaction properties between impurities and the host matrix can be determined by the response of superconducting critical temperature ($T_c$) to impurity concentration. Introduction of non-magnetic impurities which can be described within Anderson model[1] does not change $T_c$ significantly. Contrary, magnetic implants can suppress $T_c$ even at dilute concentration[2] < 1% at which the impurities are considered to be non-interacting. Naturally there is also a possibility that impurities start to interact with each other ferromagnetically or antiferromagnetically (AF) at certain concentrations[3]. Antiferromagnetic interactions can manifest themselves as non-monotonic behavior of $T_c$ with increased impurity dose. Equation 1 describes the dependence of $T_c$ on concentration and magnetic coupling parameter $\alpha$[3,4]. In the dilute limit Equation 1 reduces to Abrikosov and Gor'kov (AG) result[2] with $\alpha = 0$.

$$\ln \frac{T_c}{T_{c0}} = \psi\left(\tfrac{1}{2}\right) - \psi\left(\tfrac{1}{2} + \tfrac{1}{4} e^{-\gamma} \frac{x}{x_c} \frac{T_{c0}}{T_c} \left[1 + \alpha \frac{x}{x_c} \frac{T_{c0}}{T_c}\right]\right)$$

**Equation 1.** In Equation 1. $T_{c0}$ is the critical temperature of the undoped superconductor, $x_c$ is the critical concentration at which superconductivity is completely suppressed, γ is Euler-Mascheroni constant and ψ is the digamma function.

In aluminum $T_c$ change is typically related to magnetic interaction of the impurity or resonant states (virtual bound states, VBS) between the impurities (e.g. Fe) and the host matrix Fermi level in the Friedel-Anderson model[5-11]. In case of resonant states, $T_c$ suppression is due to reduction of probability of a Cooper pair formation. The dependence of corresponding $T_c$ suppression rate on impurity concentration has weaker dependence compared to the magnetic pair-breaking mechanism[5-15]. One exception is non-magnetic, but VBS/spin fluctuating Mn in Al,[5,10-13,16-21] which gives an initial $T_c$ suppression rate even stronger than magnetic Gd in La[22]. It is rather tempting to find out whether Fe and Mn impurities can provide distinct $T_c$ suppression in thin Al films compared to the bulk material in the limit of dilute impurity concentration. AG theory can be used as a reference as the model is flexible enough to take into account various types of magnetic interactions[23].

Aluminum samples were fabricated on $SiO_2$/Si and SiN/Si substrates by electron beam lithography on PMMA mask followed by a lift-off process. After patterning the substrates, Al was e-beam deposited in ultra-high vacuum (UHV) conditions. The samples were shaped as long wires with length 2 mm, width 50 µm, and thickness varying from 25 nm to 250 nm. Structures were implanted with $Mn^{+q}$ and $Fe^{+q}$ ions (q = 6 - 13) using UHV-chamber equipped with electron cyclotron resonance ion source ECRIS 6.4 GHz. Homogeneity of the ion implanted impurity distribution (and therefore reduction of superconducting transition width) can be improved by implantation with variable ion energy. Thinner aluminum films were implanted with single energy of 12 keV (20 nm AlFe) and 25 keV (42 nm Mn and 55 nm Fe) giving peak concentration close to half-thickness of the film. Thicker films were implanted with varying energies of 20/60 keV (72 nm AlMn), 35/80 keV (95 nm and 115 nm AlFe) and 80/225 keV (250 nm film). For example, full coverage of Fe over a 250 nm thick Al film is not possible with one implantation energy. The resulting distribution to some extend can be smoothed by temperature activated diffusion, but extra thermal treatment causes additional complications and was not used in present experiments. Instead, implantation with two energies of 80 keV and 225 keV was used resulting in relatively even impurity distribution shown in Fig.1.

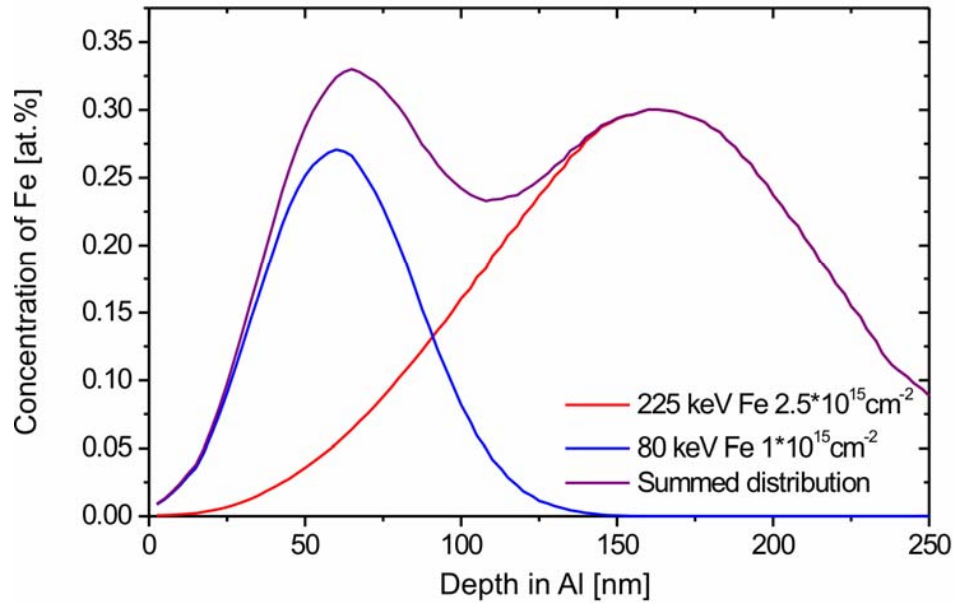

**Figure 1**. SRIM2003 code[24] is used for estimation of resulting depth distribution of Fe in Al when two implantation energies are used. Combined dose is $3.5*10^{15}$ $1/cm^2$ resulting in ~0.3 at.% quasi-homogeneous distribution of impurities from 50 nm to 200 nm.

Additionally depth distribution is typically flattened with high implantation doses due to radiation enhanced diffusion and sputtering resulting in impurity migration deeper into the target. In our case, Al film remains polycrystalline in spite of high Fe or Mn dose eventually because Al is efficiently annealed already at room temperatures. No signs of amorphization are observed except, probably, the very thin layer at the top of the film (Fig. 3a). However, in this region chemical surface damage is already inevitable.

To reach required impurity concentrations the samples were implanted with different fluencies from $10^{14}$ $cm^{-2}$ to $10^{17}$ $cm^{-2}$. During implantations the ion beam current was kept at a level of 100 nA/$cm^2$ to avoid overheating of the samples. Films thickness was measured with Tectra P-15 surface profiler, and it was found to remain almost unchanged after implantations presumably due to low sputtering rate of the natural oxide on Al surface. Minor deviations towards slight broadening of the calculated impurity depth distributions were anticipated because of the radiation enhanced diffusion induced by the ion beam.

Directly pumped $^4$He bath cryostat was used to cool Al films to the superconducting state. The temperature range of the cryostat can be varied from 4.2 K down to 0.95 K with a temperature stability of about 0.1 mK. The critical transition temperature $T_c$ from the normal to superconducting state was measured by the four probe method (Fig.2). For lower $T_c$ measurements on heavy implanted structures self-made $^3$He$^4$He dilution refrigerator capable of producing sub-100 mK temperatures was used.

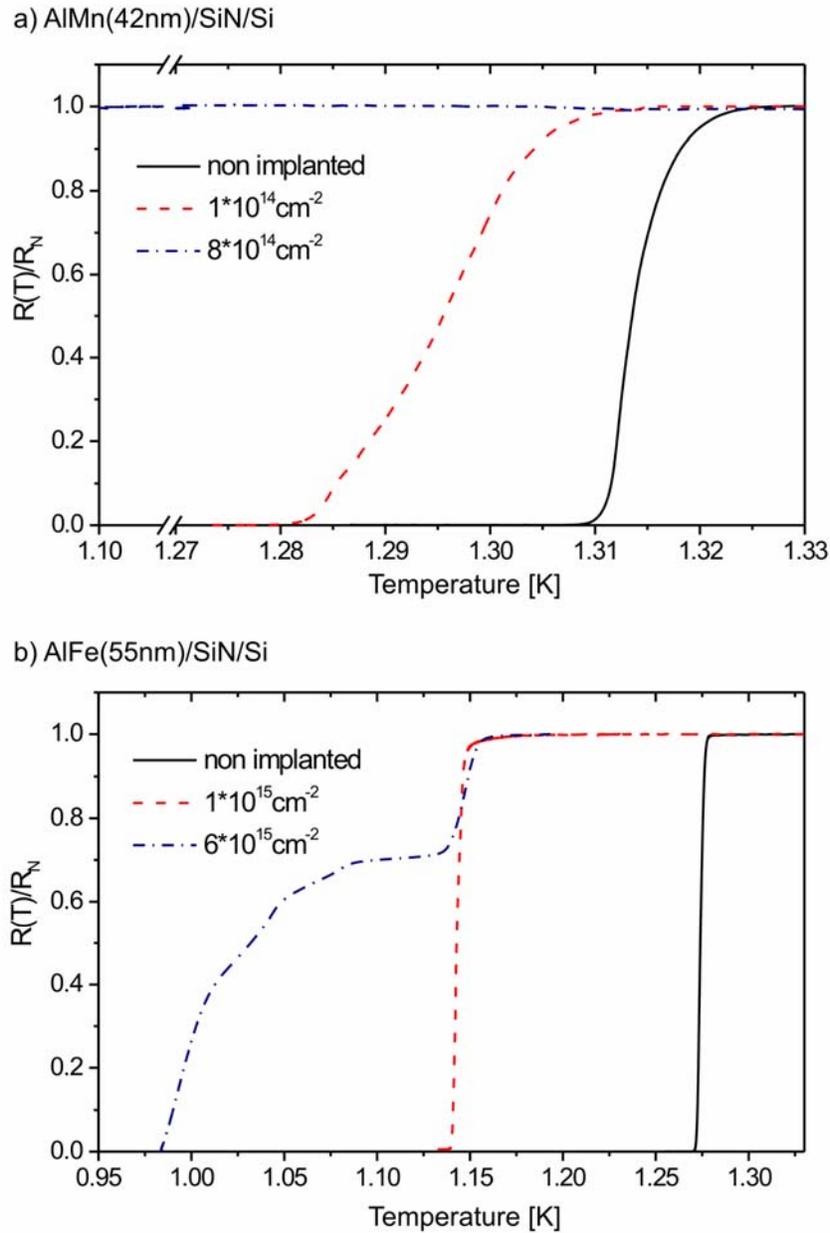

**Figure 2.** a) Temperature dependence of the normalized resistance $R(T)/R_N$ for non implanted and Mn implanted Al 42 nm thick films on SiN/Si substrate. b) Same data for non implanted and Fe implanted Al 55 nm thick films on SiN/Si substrate. The wide $R(T)/R_N$ transition in case of dose $6*10^{15}$ cm$^{-2}$ is presumably due to inhomogeneous Fe distribution in Al matrix.

Before implantation the transition temperatures turned out to fall into two groups corresponding to different substrates. Non-implanted Al films grown on $SiO_2$/Si and SiN/Si have critical temperature at around 1.40 K and 1.27 K, respectively. Difference of $T_c$ of non-implanted samples of the same thickness was within the detection level of ± 0.005 K. Therefore inevitable small variations of the initial critical parameters between samples were not an issue as $T_c$ changes of the order of 0.5 K - 1 K were finally observed. We performed progressive ion implantation alternated with $T_c$ measurements tracing the evolution of the sample critical parameters eliminating mentioned uncertainty due to initial $T_c$ variation.

Samples were analyzed using high resolution electron microscopy (HREM) revealing polycrystalline Al structure with an average grain size ~ 40 nm (Fig.3a). No deformation of the Al crystal lattice has been observed. As all structures were prepared in UHV using 99.999% pure Al, it is very unlikely that the difference in the initial critical temperatures is due to residual impurities left in the samples after fabrication. The residual impurity concentration should be several orders of magnitude lower compared to the smallest dose of magnetic implants causing detectable changes of $T_c$. The difference in the initial transition temperature is presumably due to mismatch of thermal expansion constants between $SiO_2$ and SiN substrates leading to mechanical stress of the thin film while cooling down to cryogenic temperatures.

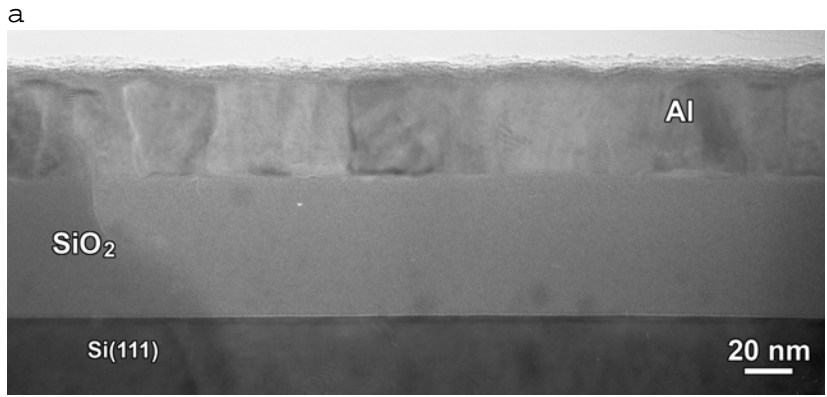

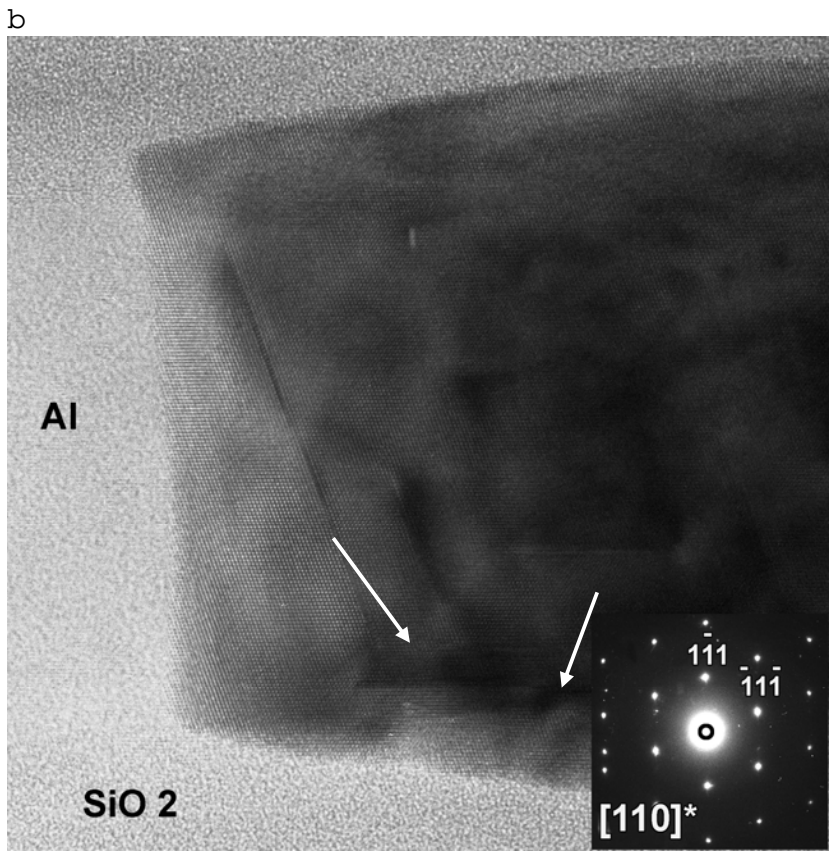

**Figure 3**. a) Low resolution TEM image of a typical structure cross section. The sample has been implanted with Fe with dose $1.2*10^{17}$ $1/cm^2$. One can clearly resolve individual crystal grains of Al on top of $SiO_2$/Si(111)substrate. b) High resolution electron microscopy image along the [110] zone of an individual Al grain of a structure implanted with Mn (42 nm thick Al film) with dose $5*10^{14}$ $1/cm^2$. Dislocations and associated stacking faults along {111} planes are marked by white arrows. Inset: electron diffraction pattern taken in [110] direction reveals undeformed cubic Fm-3m lattice for the Al grain. Note absence of any inclusions in HREM and distortion of ED pattern due to Mn implantation.

Implantation of Mn into Al films grown on SiN/Si substrates leads to a substantially strong decrease of $T_c$ than the same concentration of implanted Fe. Apart from the dose and type of magnetic implants, the $T_c$ reduction also depends on the film thickness (Fig. 4). We believe that for a given sample the resulting $\Delta T_c$ is determined by several competing contributions: magnetic ion concentration, radiation-induced disorder and dynamic annealing of the dislocations. The last mechanism is known to be rather effective in aluminium at room temperatures. Though we do not see any explicit evidence of radiation-induced disorder in our samples using TEM analysis, we cannot exclude contribution of this mechanism to $T_c$ modification. In the first approximation, for a given material the radiation damage should depend on dose and energy of implants. To obtain the same homogeneous concentration of impurities thicker films should be implanted with higher energies than the thinner ones (Fig. 1). Hence, the resulting $T_c$ might have a rather non-trivial dependence on film thickness and magnetic ion dose (Fig. 1). While the general trend is clear: the higher the dose, the lower the critical temperature.

The mean critical temperature $T_c$ was defined to be the half of the normal state resistance $R_N$, i.e. $R(T_c) = \frac{1}{2}R_N$ in Fig.4. Error bars are equivalent to the R(T) transition half-width (Fig.2), except for the points of 42 nm Mn. Here the low temperature limit of cryostat ~0.95 K was reached at ~0.1% Mn concentration, setting an upper limit of $T_c$ for 42 nm (Fig.4b) Al films. Estimation is needed because $^3He^4He$ dilution refrigerator capable of going down to the mK range was not available during the 42nm Al/Mn film measurements. The $T_c$ value for ~0.1% Mn concentration is expected to be within the range from 0.45 K [25] to 0.95 K that is shown as a large error bar in Fig. 4b.

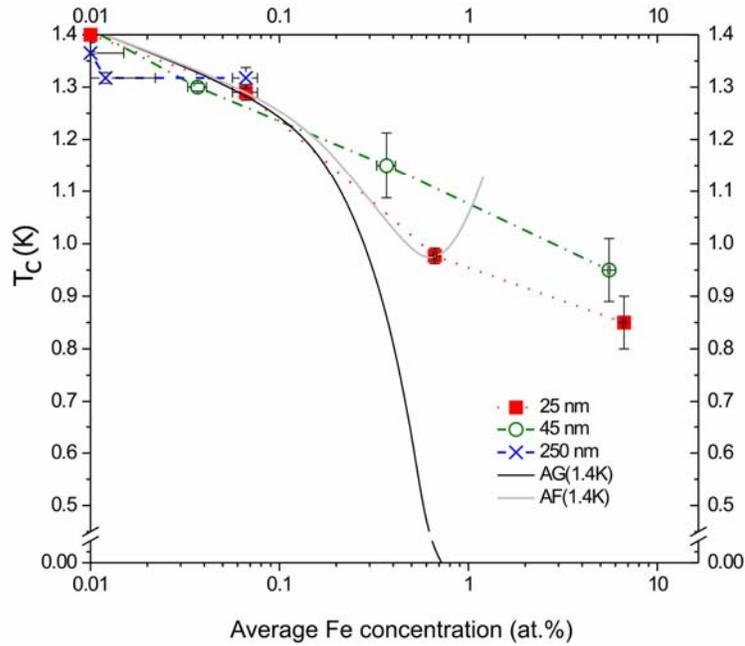

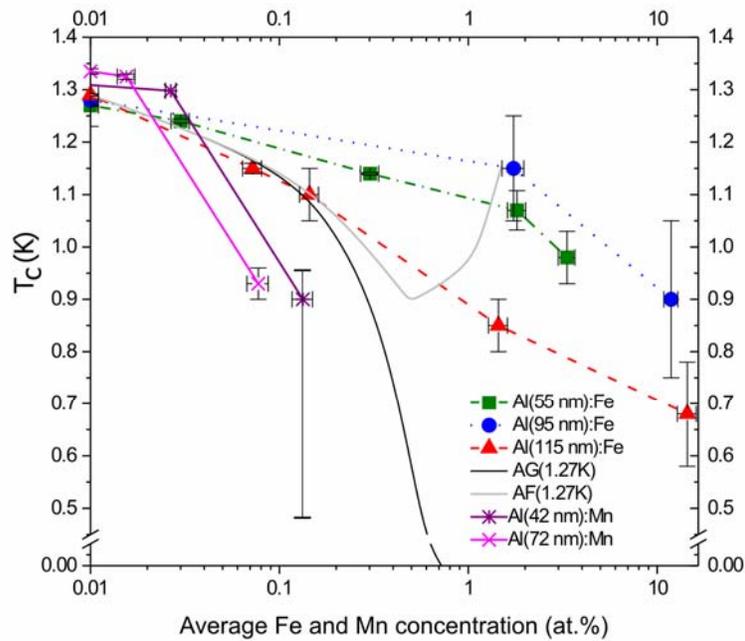

**Figure 4.** a) Dependence of critical temperature in Al films grown on $SiO_2$/Si substrates on concentration of Fe dopants. b) Same data for films grown on SiN/Si substrates and subsequently implanted with Fe and Mn. Lines stand for theoretical predictions of the Abrikosov- Gor'kov (AG) model and the one assuming impurity atoms interacting antiferromagnetically (AF).

In the dilute limit of low magnetic impurity concentration suppression of superconductivity in Al films by Fe is qualitatively consistent with the AG model. The model predicts a linear correspondence between impurity concentration and $T_c$ change. Critical temperature suppression depends on both the impurity concentration and its interaction strength in the host matrix. The $T_c$ versus concentration dependence predicted by the AG model assuming magnetic interactions between Fe and Al atoms (Eq.1 with α = 0) is presented in Fig.4 (solid black lines). According to the model, there should be a critical concentration at which transition to the superconducting state is no longer possible. By applying the AG model to our experimental data the critical concentration of Fe in Al was estimated to be around 0.72 at.%. In reality, however, with increasing Fe concentration interactions between the impurity atoms start to occur. This may lead to either an increase or decrease of $T_c$ with increasing impurity concentration that would make superconducting behavior of Al films more complex. Assuming AF interacting Fe atoms (Eq.1 with α < 0), the $T_c$ concentration dependence was calculated and shown in Fig.4 by gray continuous lines. It is obvious that experimental $T_c$ behavior follows neither the AG, nor AF models. AG theory predicts an overly strong response to the impurity concentration in the dilute limit, whereas at high concentrations there are no reasonable values of the antiferromagnetic coupling parameter α in Eq.1 that would allow even qualitative fitting of the experiment. Ferromagnetic ordering (α > 0) is not considered here because it would imply even stronger suppression rate than in the AG model that is not found in experiments.

Data from Mn implantation in Al films seems to be insufficient for making a decisive conclusion on applicability of AG model. A comparison to known experiments shows that $T_c$ is suppressed at a faster rate with Mn than with Fe[7,25,26]. Experiments[26] showed that ion implantation with Mn is capable of strong $T_c$ suppression ~15% in a 60 nm thick film at 0.05 at.% concentration. Corresponding results have been obtained in present work for Mn implantation of 72 nm thick Al film (Fig.4b). On the other hand, a fast suppression down to 50 mK with 0.3 at.% Mn concentration has been measured for sputter deposited AlMn films[27]. Similar conclusion can be deduced from our own experimental data by straightforward extrapolation of 72 nm Al/Mn suppression rate for higher concentrations of Mn impurities (Fig. 4b).

It is known that transition metals do not exhibit a localized permanent magnetic moment when incorporated into an Al matrix. However, doping with transition elements was observed to result in significant $T_c$ suppression in bulk Al[7]. From the theoretical study by Friedel and Anderson[8,10] it follows that mixing of resonant d-states of Fe incorporated into Al matrix with Al Fermi states is responsible for $T_c$ suppression, and this seems to be consistent with experiments of Boato and Ruggiero[7,12,25,26]. The suppression mechanism of Mn in Al is likely to have different features compared to Fe in Al because of the stronger response at low concentration.[9] However, it has been shown that Mn does not possess a permanent magnetic moment that follows Curie-Weiss law in an Al matrix making the AG model an unlikely choice to describe $T_c$ suppression.[20] Classification of Mn properties in Al has been controversial because it belongs to the VBS model due to its transition metal properties, but the strong $T_c$ suppression effect and XPS measurements[28] imply that localized spin exists at least over a short time scale but interpretation of these results are questioned by more resent theoretical calculations.[20] In neutron diffraction studies it has been shown that Mn possesses a magnetic moment but it is compensated by surrounding antiferromagnetic electron cloud at low temperatures. Therefore, Mn in Al matrix is considered to be a spin fluctuating system[6] belonging to the VBS model.

One cannot exclude that ion implantation alters the inter-atomic distance in Al matrix. This effect might account for the difference of properties of implanted thin films compared to bulk samples[6]. However, even for the highest implantation doses HREM electron diffraction (ED) analysis did reveal typical cubic Fm-3m lattice for Al without any traces of crystalline lattice deformation which can be attributed to ion implantation. Inevitable imperfections (Fig. 3b, white arrows) eventually do reside in Al microcrystalline grains just after the deposition of the film, and their origin is not related to ion implantation. It should be noted that neither HREM, nor ED analysis revealed any traces of impurity cluster formation in hosting Al matrix.

In summary, it has been shown that ion implantation of Fe and Mn into Al thin films can be used for effective modification of Al superconductive properties. Critical temperature of the transition to superconducting state was found to decrease gradually with implanted Fe concentration being additionally dependent on the substrate type. It was found that suppression by Mn implantation is much stronger compared to Fe. At low concentrations of implanted ions, suppression of the critical temperature can be described with reasonable accuracy by existing models AG and AF, while at concentrations above 0.1 at. % a pronounced discrepancy between the models and experiments is observed. The interaction between impurities and conduction electrons can in principle be estimated or identified from the experiments. The implanted impurities (Fe and Mn) have a different effect on $T_c$ suppression due to a distinct underlying interaction mechanism. Neither Fe, nor Mn are considered to possess a permanent magnetic moment in Al, though Mn is commonly used as an effective $T_c$ suppressor for various applications. Existing (bulk) theoretical models do not take into consideration the dependence of critical temperature suppression on film thickness, which is clear from our results. However, one cannot conclude that this dependence comes solely from the response of a 2D system to magnetic ion implantation. Apart from concentration and type of magnetic impurities, critical temperature of a superconducting film depends on other parameters: substrate and film thickness. Critical temperature thickness dependence of undoped thin superconducting films is known for decades, while commonly accepted explanation of the phenomenon is still absent. Nevertheless for given aluminium film one can trace a general trend: the higher the concentration of magnetic implants, the lower the critical temperature. It has been shown that the ion implantation method enables systematic study of evolution of low dimensional system properties with respect to the implanted ion dose. In this way, artifacts due to variation in specimens (which is an important factor particularly when a system dimensions are scaled down) can be avoided. Accurate results can be obtained and delicate $T_c$ changes can be observed reliably.

Authors would like to acknowledge Dr. Ludo Rossou for excellent TEM specimen preparation. The work was supported by the Academy of Finland under the Finnish Centre of Excellence Program 2000-2005 No. 44875, the Nuclear and Condensed Matter Program at JYFL; the Russian Federation for Basic Research (Grant No. 04-02-17397-A)" Experimental study of spin-polarized non-equilibrium quasiparticle excitations in a superconductor;" and the EU Commission FP6 NMP-3 Project No. 505587-1 "Superconductivity-Ferromagnetism Interplay in Nanostructured Systems" (SFINX).